\documentclass{elsart}

 \usepackage{graphicx}

\usepackage{amssymb}

\begin{document}

\begin{frontmatter}



\title{First-order phase transition of the tethered membrane model on spherical surfaces}

\author[label1]{Isao Endo} and
\author[label2]{Hiroshi Koibuchi}
\ead{koibuchi@mech.ibaraki-ct.ac.jp}

\address[label1]{Department of Electrical and Electronic System Engineering, Ibaraki College of Technology, 
Nakane 866, Hitachinaka,  Ibaraki 312-8508, Japan}

\address[label2]{Department of Mechanical and Systems Engineering, Ibaraki College of Technology, 
Nakane 866, Hitachinaka,  Ibaraki 312-8508, Japan}

\begin{abstract}
We found that three types of tethered surface model undergo a first-order phase transition between the smooth and the crumpled phase. The first and the third are discrete models of Helfrich, Polyakov, and Kleinert, and the second is that of Nambu and Goto. These are curvature models for biological membranes including artificial vesicles. The results obtained in this paper indicate that the first-order phase transition is universal in the sense that the order of the transition is independent of discretization of the Hamiltonian for the tethered surface model. 
 \end{abstract}

\end{frontmatter}


\section{Introduction}\label{intro}
A considerable number of attempts have been made at clarifying the phase structure of the surface model of Helfrich, Polyakov, and Kleinert \cite{HELFRICH-1973,POLYAKOV-NPB1986,NELSON-SMMS2004,David-TDQGRS-1989,David-SMMS2004,Wiese-PTCP2000,Bowick-PREP2001,WHEATER-JP1994,Peliti-Leibler-PRL1985,David-EPL1986,DavidGuitter-EPL1988,PKN-PRL1988,BKS-PLA2000,BK-PRB2001};  henceforce we will refer to this as the HPK model. Tethered surface models are defined on triangulated fixed connectivity surfaces representing polymerized biological membranes or membranes in the gel phase \cite{Bowick-PREP2001}, and they are classified into a major class of the HPK model \cite{KANTOR-NELSON-PRA1987,KANTOR-SMMS2004,WHEATER-NPB1996,BCFTA-JP96-NPB9697,KY-IJMPC2000-2,KOIB-PLA-2003-2,KOIB-PRE-2004-2,KOIB-PLA-2005-1,KOIB-PLA-2005-2,KD-PRE2002,KOIB-PRE-2004-1,KOIB-PRE-2005-1}. Fluid surface models are considered a different class of the HPK model defined on dynamically triangulated surfaces representing these biological membranes in the fluid phase, however, we will not discuss the fluid surface model in this paper \cite{CATTERALL-NPBSUP1991,AMBJORN-NPB1993,ABGFHHM-PLB1993,BCHHM-NPB9393,KY-IJMPC2000-1,KOIB-PLA200234,KOIB-EPJB-2004,KOIB-EPJB-2005}. 

An interesting problem that remains unanswered at present is whether the tethered self-intersecting (or phantom) surface model undergoes a continuous transition or a discontinuous one. The model is expected to undergo a phase transition separating the smooth phase in the limit $b\!\to\! \infty$ from the crumpled one in the limit $b\!\to\! 0$. The mean field analysis predicts that the model undergoes a discontinuous phase transition \cite{PKN-PRL1988}, whereas the large-$D$ expansion predicts a continuous transition \cite{DavidGuitter-EPL1988}. Biological membranes including artificial vesicles were thought to undergo the crumpling transition. However, we currently have no experimental evidence for such a transition in biological membranes, except for an investigation on artificial vesicles \cite{Chaieb-2005-Benasque}. 

In a recent numerical study, it was reported that the tethered model of HPK undergoes a first-order phase transition on spherical surfaces relatively larger than those used so far \cite{KOIB-PRE-2005-1}. The bending energy and the Gaussian term used in \cite{KOIB-PRE-2005-1} are those widely accepted as standard discrete Hamiltonians in numerical studies on the HPK model. The bending energy is of the form $1\!-\!{\bf n}_i\cdot {\bf n}_j$,  where ${\bf n}_i$ is the unit normal vector of the triangle $i$. The Gaussian term is given by the sum of bond length squares. 

However, it has not yet been clarified whether the order of phase transition depends on the discretization of the Hamiltonian. If the order of phase transition changes depending on the choice of the discretization, the phase transition might be seen only in the lattice models, and as a consequence we can obtain no information on the phase transition of the continuous model by using the numerical simulations. Hence, it is natural to consider that the order of the transition depends on the discretization of Hamiltonian. Therefore, it is worthwhile to investigate further the tethered model defined by Hamiltonian which is different from the above-mentioned bending energy and Gaussian term.

In this paper, we studied three types of tethered models on triangulated spherical surfaces, which are obtained by dividing the icosahedron as in \cite{KOIB-PRE-2005-1}. The Hamiltonian of the first model contains the ordinary Gaussian energy and a bending energy, which is defined by the normal vector of vertices. Hence, the discretization of the bending energy in the first model is slightly different from that of the standard bending energy. It was previously reported that the first model undergoes a discontinuous transition on Voronoi lattices \cite{KOIB-PRE-2004-1}. However, it remains unclear whether the discontinuous transition can be seen on any other lattices. Thus, we can also check that the discontinuous transition is independent of the lattice structure by investigating the first model defined on lattices constructed from the icosahedron. 

The second model is called the Nambu-Goto surface model \cite{KY-IJMPC2000-2,KOIB-PRE-2004-2,KY-IJMPC2000-1}. At present, it is unclear whether the Nambu-Goto model undergoes a discontinuous phase transition when the model is defined on surfaces with extrinsic curvature. The area energy, which is given by the sum of the area of the triangles, corresponds to the Gaussian energy in the first (HPK) model. It is well-known that the Nambu-Goto model without the bending energy term is not well-defined due to the lack of the smooth phase and the appearance of spikes on the surface \cite{ADF-NPB-1985}. The area term imposes a constraint only on the area of the triangles, and hence allows the appearance of spiky configurations made up of oblong triangles. On the contrary, some additional terms in the Hamiltonian can make the model well-defined. In fact, it was reported recently that the Nambu-Goto model with a deficit angle term, which is an intrinsic curvature, is well-defined and undergoes a discontinuous transition between the smooth phase and a tubular phase \cite{KOIB-PRE-2004-2}. The second model in this paper is also well-defined \cite{KY-IJMPC2000-1,KY-IJMPC2000-2}, because the Hamiltonian includes a bending energy, which is an extrinsic curvature defined according to the dual lattice formulation of the discrete mechanics by Lee \cite{TDLEE}. 

The third model is a tensionless model whose Hamiltonian contains the standard bending energy and a hard-wall potential which gives an upper bound on the bond length. There is no Gaussian term in the Hamiltonian. The bending energy of the model is identical to the one in \cite{KOIB-PRE-2005-1}. Therefore, the discretization of the third model is the same as in \cite{KOIB-PRE-2005-1}. However, it is not yet clear whether a tensionless tethered model undergoes a discontinuous transition.

It is nontrivial whether those three discrete models lead to the same result on the order of the transition. The standard bending energy for the HPK model is not equivalent to the above mentioned bending energy for the Nambu-Goto surface model. In fact, the Nambu-Goto model with the standard bending energy is ill-defined because of the lack of the smooth phase and the appearance of spikes in the whole region of the bending rigidity \cite{ADF-NPB-1985}. Thus, it is worthwhile to study numerically whether these three models have the same continuous limit for the order of the transition.  

We will see that the three models undergo a discontinuous transition on spherical surfaces relatively larger than those used to date, and then we will understand that the discontinuous transition of the model is independent of the discretization of the Hamiltonian. Therefore, the results obtained in this paper together with those in \cite{KOIB-PRE-2005-1} lead us to conclude that a first-order phase transition can be observed in a spherical tethered surface model defined by Hamiltonian that includes a bending energy term. This means that the first-order transition of triangulated surfaces is universal in the sense that the order of the transition is independent of discretization of the Hamiltonian.  

\section{Models}
The partition function of the models is defined by
\begin{equation}
\label{part-func}
Z = \int \prod_{i=1}^N d X_i \exp\left[ -S(X) \right],
\end{equation}
where $N$ is the total number of vertices, $S(X)$ is Hamiltonian of the model. The center of the surface is fixed to remove the translational zero-mode. The self-avoiding interaction is not assumed in the Hamiltonian, hence the surface is considered to be phantom.

We investigate three types of model which differ from each other in energy terms included in the Hamiltonian $S(X)$. The first model denoted by {\it model 1} is defined by
\begin{eqnarray}
\label{S1S2-model-1}
 && S(X)=S_1 + b S_2, \nonumber \\
 &&S_1=\sum_{(i,j)} \left(X_i-X_j\right)^2, \quad S_2=\sum_{i=1}^{N}\sum_{j(i)}\left[1-{\bf n}(i)\cdot{\bf n}_{j(i)} \right],  \quad ({\rm model \;1})
\end{eqnarray}
where $b$ is the bending rigidity, ${\bf n}_{j(i)}$ is the unit normal of the triangle $j(i)$ meeting at the vertex $i$. 
The symbol ${\bf n}(i)$ in Eq. (\ref{S1S2-model-1}) is a normal vector of the vertex $i$, and it is defined by
\begin{equation}
\label{normal}
{\bf n}(i)= {{\bf N}_i \over \vert {\bf N}_i\vert}, \quad {\bf N}_i = \sum _{j(i)} {\bf  n}_{j(i)} A_{\it \Delta_{j(i)}},
\end{equation}
where $\sum _{j(i)}$ denotes the summation over triangles  $j(i)$ linked to the vertex $i$, and  $A_{\it \Delta_{j(i)}}$ is the area of $j(i)$. Model 1 is considered a HPK model as mentioned in the introduction.  

The second model denoted by {\it model 2} is defined by
\begin{eqnarray}
\label{S1S2-model-2}
 && S(X)=S_1 + b S_2, \nonumber \\
 &&S_1=\sum_{\it \Delta} A_{\it \Delta}, \quad S_2 = \sum _{\it \Delta} {1\over A_{\it \Delta} } \sum_{i=1}^3 l_i^2 \left(1 - \cos \theta_i\right),  \quad ({\rm model \;2})
\end{eqnarray}
where $A_{\it \Delta}$ is the area of the triangle ${\it \Delta}$, $l_i$ is the edge length of ${\it \Delta}$, and $\theta_i$ is the angle between the triangles attaching to the edge $l_i$.

\begin{figure}[hbt]
\centering
\includegraphics[width=8.5cm]{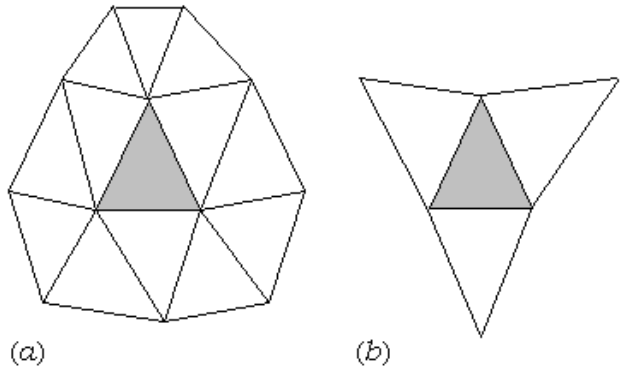}
 \caption{The range of interaction of the shaded triangle in (a) model 1, and (b) model 2. $A_{\it \Delta}$ in Eq. (\ref{S1S2-model-2}) is the area of the shaded traiangle in (b), and $l_i$ is the edge length of the triangle.}
\label{fig-1}
\end{figure}
Figures \ref{fig-1}(a), (b) show the triangles whose normal vectors are interacting with one of the shaded triangles corresponding to model 1 and model 2, respectively.  $A_{\it \Delta}$ in Eq. (\ref{S1S2-model-2}) is the area of the shaded triangle in Fig. \ref{fig-1}(b), and $l_i$ in Eq. (\ref{S1S2-model-2}) is the edge length of the triangle. 

 It should be noted that the Nambu-Goto model is considered a classical analogue of the Polyakov string model. In fact, the Polyakov string model reduces to the Nambu-Goto model if the metric, which is a dynamical variable of the Polyakov string, is fixed to the induced metric \cite{GSW-SST-1987-book-1}.

The third model denoted by {\it model 3} is defined by the Hamiltonian
\begin{eqnarray}
\label{S2V-model-3}
&& S(X)=b S_2 + V, \nonumber \\
&&S_2=\sum_{i,j}\left(1-{\bf n}_i\cdot{\bf n}_j \right), \quad V=\sum_{(ij)} V(|X_i-X_j|), \quad ({\rm model \;3}). 
\end{eqnarray}
$S_2$ in $S(X)$ of Eq. (\ref{S2V-model-3}) is the bending energy, which was already introduced in the Introduction, and $V$ is a hard-wall potential, which gives an upper bound on the bond length. $\sum_{(i,j)}$ in $V$ denotes the sum over bonds $(i,j)$ connecting the vertices $i$ and $j$, and $\sum_{i,j}$ in $S_2$ is the sum over triangles $i,j$ sharing a common bond. The symbol ${\bf n}_i$ in $S_2$ denotes the unit normal vector of the triangle $i$. The range of interaction of ${\bf n}_i$ in $S_2$ is identical to that of model 2, shown in Fig. \ref{fig-1}(b). The symbol $V(|X_i\!-\!X_j|)$ in Eq. (\ref{S2V-model-3}) is the potential between the vertices $i$ and $j$, and is defined by
\begin{equation}
\label{V}
V(|X_i-X_j|)= \left\{
       \begin{array}{@{\,}ll}
       0 & \quad (0< |X_i-X_j| < r_0), \\
      \infty & \quad ({\rm otherwise}). 
       \end{array}
       \right. 
\end{equation}
The value of $r_0$ in the right hand side of Eq. (\ref{V}) is fixed to $r_0\!=\!\sqrt{1.1}$. Then we have $\langle \sum (X_i\!-\!X_j)^2 \rangle /N \simeq 3/2$, which is satisfied when the Gaussian term $S_1\!=\!\sum (X_i\!-\!X_j)^2$ is included in the Hamiltonian without the hard-wall potential $V$. 

If it were not for the constraint $|X_i\!-\!X_j| \!<\! r_0$, the size of the surface grows larger and larger in the MC simulations. Thus the constraint $|X_i\!-\!X_j| \!<\! r_0$ is necessary for the bond length to have a well-defined value when the Gaussian term $S_1$ is not included in the Hamiltonian. We note that the model seems to depend on a hidden length scale introduced by $r_0$, and as a consequence the model appears to be ill-defined. However, we checked in Ref. \cite{KOIB-PLA-2003-2} that there is no $r_0$-dependence on the results. This is a consequence of the scale invariant property of the model, which will be mentioned in the next section. Therefore, we use $r_0\!=\!\sqrt{1.1}$ in the MC simulations.

We should comment on a relationship between the discrete bending energy $S_2$ in Eqs. (\ref{S1S2-model-1}), (\ref{S1S2-model-2}), and (\ref{S2V-model-3})  and the continuous one of the Polyakov-Kleinert string model. The functional action of the Polyakov-Kleinert string in ${\bf R}^3$ is defined by
\begin{equation}
\label{CONT-S1S2}
S(X,g) =  {a \over 2}\int_M \sqrt {g} \,d^2x \, g^{ab} {\partial X^\mu \over \partial x^a} {\partial X^\mu \over \partial x^b} + {b \over 2} \int_M \sqrt {g} \, d^2x  K_a^b K_b^a,
\end{equation}
where $(g_{ab})$ is the first fundamental form (or the metric tensor) of the surface $M$. The $K_a^b$ is defined by $K_a^b \!=\! g^{bc} K_{ac}$, where $K_{ab}$ is the second fundamental form (or the extrinsic curvature tensor) on $M$ in ${\bf R}^3$ \cite{David-SMMS2004}. In $S(X,g)$, $X$ denotes a mapping from the surface $M$ to ${\bf R}^3$, where the image $X(M)$ is the membrane.  

It is easy to see that $(1/2) \int_M \sqrt {g} \, d^2x  K_a^b K_b^a$ in the second term of $S(X,g)$ in Eq. (\ref{CONT-S1S2}) is equivalent to $S_2$ in Eqs. (\ref{S1S2-model-1}), (\ref{S1S2-model-2}), and (\ref{S2V-model-3}) if the metric tensor $(g_{ab})$ is given by the induced one; $g_{ab} \!=\! {\partial X^\mu \over \partial x^a} {\partial X^\mu \over \partial x^b}$. In fact, the second fundamental form $K_{ab}$ can be expressed by $K_{ab}\!=\!n^\mu\partial_a \partial_b X^\mu\!=\!-\partial_a n^\mu\partial_b X^\mu$, where $n^\mu$ is the unit normal vector of the surface. Then by using this relation and the induced metric, the second term in $S(X,g)$ can be expressed by $(1/2) \int_M \sqrt {g} \, d^2x g^{ab} \partial_a n^\mu \partial_b n^\mu $. This action is identical to the one of the $\sigma$-model, and hence a natural discretization of this term is given by $S_2$ in Eqs. (\ref{S1S2-model-1}), (\ref{S1S2-model-2}), and (\ref{S2V-model-3}).

\section{Monte Carlo technique}
The canonical Monte Carlo (MC) technique is used to update the variable $X$. The new position $X^\prime_i$ of the vertex $i$ is given by $X^\prime_i\!=\!X_i\!+\!{\it \Delta} X$, where ${\it \Delta} X$ is chosen randomly in a small sphere for model 1 and model 2. The radius of the small sphere is chosen at the start of the MC simulations to maintain about $50\%$ acceptance rate. The new position $X^\prime_i$ is accepted with the probability ${\rm Min}[1,\exp\left(-{\it \Delta}S\right) ]$, where ${\it \Delta}S$ is given by ${\it \Delta}S\!=\!S({\rm new})\!-\!S({\rm old})$.

Under the constraint of Eq.(\ref{V}) for the edge length for model 3, the new position $X^\prime_i$ is accepted with the probability ${\rm Min}[1,\exp\left(-{\it \Delta}S\right) ]$; namely the Metropolis accept/reject procedure is applied only for $X^\prime_i$ satisfying the constraint of  Eq.(\ref{V}). This constraint for the edge length is imposed by $r_0^2\!=\!1.1$ as described in the previous section. The rate of acceptance for the constraint of Eq. (\ref{V}) is about $50\%$.  

A random number called Mersenne Twister \cite{Matsumoto-Nishimura-1998} is used in the MC simulations. We use two-sequences of random numbers; one for the 3-dimensional move of vertices $X$ and the other for the Metropolis accept/reject in the update of $X$.

For model 1 and model 2, the minimum bond length is not assumed, while the minimum area of triangle is assumed to be $10^{-7}\times A_0$, where $A_0$ is the mean area of triangles computed at every 250 MCS (Monte Carlo sweeps)  and is almost constant throughout the MC simulations. The area of almost all triangles generated in the simulations is larger than the lower bound $10^{-7}\times A_0$.

 For model 3, the minimum area of triangle is assumed to be $1\!\times\!10^{-7}$. The area of almost all triangles generated in the MC simulations is larger than the lower bound $1\!\times\!10^{-7}$. The minimum bond length is assumed to be $1\!\times\!10^{-7}$, however, we find no bond whose length is less than the lower bound  $1\!\times\!10^{-7}$ in the MC simulations. This means that no constraint is imposed on the lower bound of the bond length. 

The Hamiltonians in Eqs. (\ref{S1S2-model-1}), (\ref{S1S2-model-2}), and (\ref{S2V-model-3}) are defined on the surfaces which are constructed uniformly by the co-ordination number. They are obtained by dividing the icosahedron. By dividing every edge of the icosahedron into $L$-pieces of the same length, we have a triangulated lattice of size $N\!=\!10L^2\!+\!2$. These lattices are characterized by $N_5\!=\!12$ and $N_6\!=\!N\!-\!12$, where $N_q$ is the total number of vertices with co-ordination number $q$; thus we have lattices in which 12 vertices are of $q_i\!=\!5$, and all other vertices $q_i\!=\!6$.

We comment on the unit of physical quantities in the model. The scale of length in the model can be arbitrarily chosen because of the scale invariant property of the partition function in Eq. (\ref{part-func}). Then, by letting $a$ be a length unit in the model, we can express all quantities with unit of length in terms of $a$. Hence, the unit of $S_1$ is $a^2$. Let $\lambda$ be the surface tension coefficient, then $S$ in Eqs. (\ref{S1S2-model-1}), (\ref{S1S2-model-2}) for model 1, model 2 can be written as $S\!=\!\lambda S_1\!+\! b S_2$. 

Thus, the unit of $\lambda$ can be written as $kT/a^2$, where $k$ is the Boltzmann constant, $T$ is the temperature. The unit of $b$ is therefore expressed by $kT$. The coefficient $\lambda$ is assumed to be $\lambda\!=\!1$ in this paper. This is always possible because of the scale invariant property of the partition function of the model. The unit of $\lambda$ can be influenced by the length unit $a$ of the model, and the value of $\lambda$ varies depending on $a$.

 However, $\lambda$ can always be fixed to $\lambda\!=\!1$. This is the reason why the length unit $a$ is arbitrary chosen in the model. Note also that varying the temperature $T$ is effectively identical to varying $b$ in the model. The dependence of $\lambda$ on $T$ can be absorbed into a redefinition of $\lambda$, which can be fixed to $\lambda\!=\!1$ due to the scale invariant property as stated above. The parameter $r_0$ in Eq. (\ref{V}) plays a role in the length unit $a$ of model 3. 

\section{Results}
First, we summarize the total number of MCS iterated in the simulations. At the transition point of model 1, $1.8\!\times\!10^8$,  $1.8\!\times\!10^8$,  and $1.5\!\times\!10^8$ MCS were iterated on the surfaces of $N\!=\!6252$, $N\!=\!4412$, and $N\!=\!2562$, respectively. At the transition point of model 2, $7.8\!\times\!10^8$,  $7\!\times\!10^8$,  and $6\!\times\!10^8$ MCS were iterated on the surfaces of $N\!=\!10242$, $N\!=\!6762$, and $N\!=\!4842$, respectively.  For model 3, the total number of MCS at the transition point was about $3\!\times\!10^8$, $5\!\times\!10^8$, and $6.5\!\times\!10^8$  for the surfaces of $N\!=\!4842$, $N\!=\!8412$, and $N\!=\!16812$, respectively. A relatively small number of MCS were performed at $b$ which is not  the transition point on each surface.

The mean square size $X^2$ is defined by
\begin{equation}
\label{X2}
X^2={1\over N} \sum_i \left( X_i -\bar X \right)^2,\qquad \bar X ={1\over N} \sum_i X_i,
\end{equation}
where ${\bar X}$ is the center of the surface.

\begin{figure}[hbt]
\centering
\includegraphics[width=8.5cm]{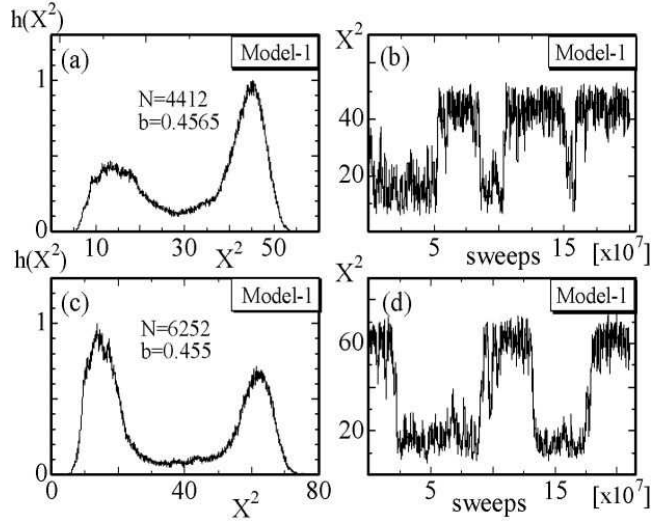}
 \caption{(a) The normalized histogram $h(X^2)$ of $X^2$ for model 1 on the $N\!=\!4412$ surface at $b\!=\!0.4565$, and (b) the variation of $X^2$ against MCS. (c) $h(X^2)$, and (d) the variation of $X^2$ obtained on the $N\!=\!6252$ surface at $b\!=\!0.455$. Two distinct peaks on each $h(X^2)$ imply that the size of surfaces in one phase is very different from that in the other phase at the transition point of model 1.  }
\label{fig-2}
\end{figure}
Figure \ref{fig-2}(a) is a normalized distribution $h(X^2)$ (or the histogram) of $X^2$ for model 1 on the surface of size $N\!=\!4412$ at the transition point $b\!=\!0.4565$. The variation of $X^2$ against MCS is plotted in Fig. \ref{fig-2}(b). Figures  \ref{fig-2}(c) and \ref{fig-2}(d) are those obtained on the surface of $N\!=\!6252$ at $b\!=\!0.455$. Two distinct peaks visible in $h(X^2)$ imply that the size of surfaces in one phase is very different from that in the other phase at the transition point. 

\begin{figure}[hbt]
\centering
\includegraphics[width=8.5cm]{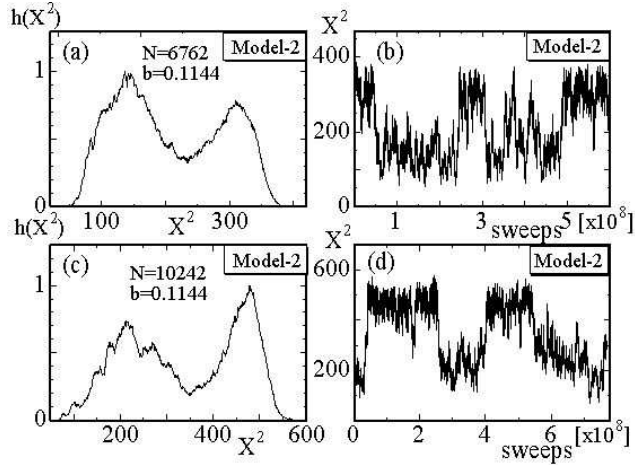}
 \caption{(a) The normalized histogram $h(X^2)$ of $X^2$ for model 2 on the $N\!=\!6762$ surface at $b\!=\!0.1144$, and (b) the variation of $X^2$ against MCS, (c) $h(X^2)$ and (d) the variation of $X^2$ obtained on the $N\!=\!10242$ surface at $b\!=\!0.1144$. }
\label{fig-3}
\end{figure}
Figure \ref{fig-3}(a) is a normalized histogram $h(X^2)$ of $X^2$ for model 2 on the surface of size $N\!=\!6762$ at the transition point $b\!=\!0.1144$. The variation of $X^2$ against MCS is plotted in Fig. \ref{fig-3}(b). Figures  \ref{fig-3}(c) and \ref{fig-3}(d) are those obtained on the surface of $N\!=\!10242$ at $b\!=\!0.1144$. 

\begin{figure}[hbt]
\centering
\includegraphics[width=8.5cm]{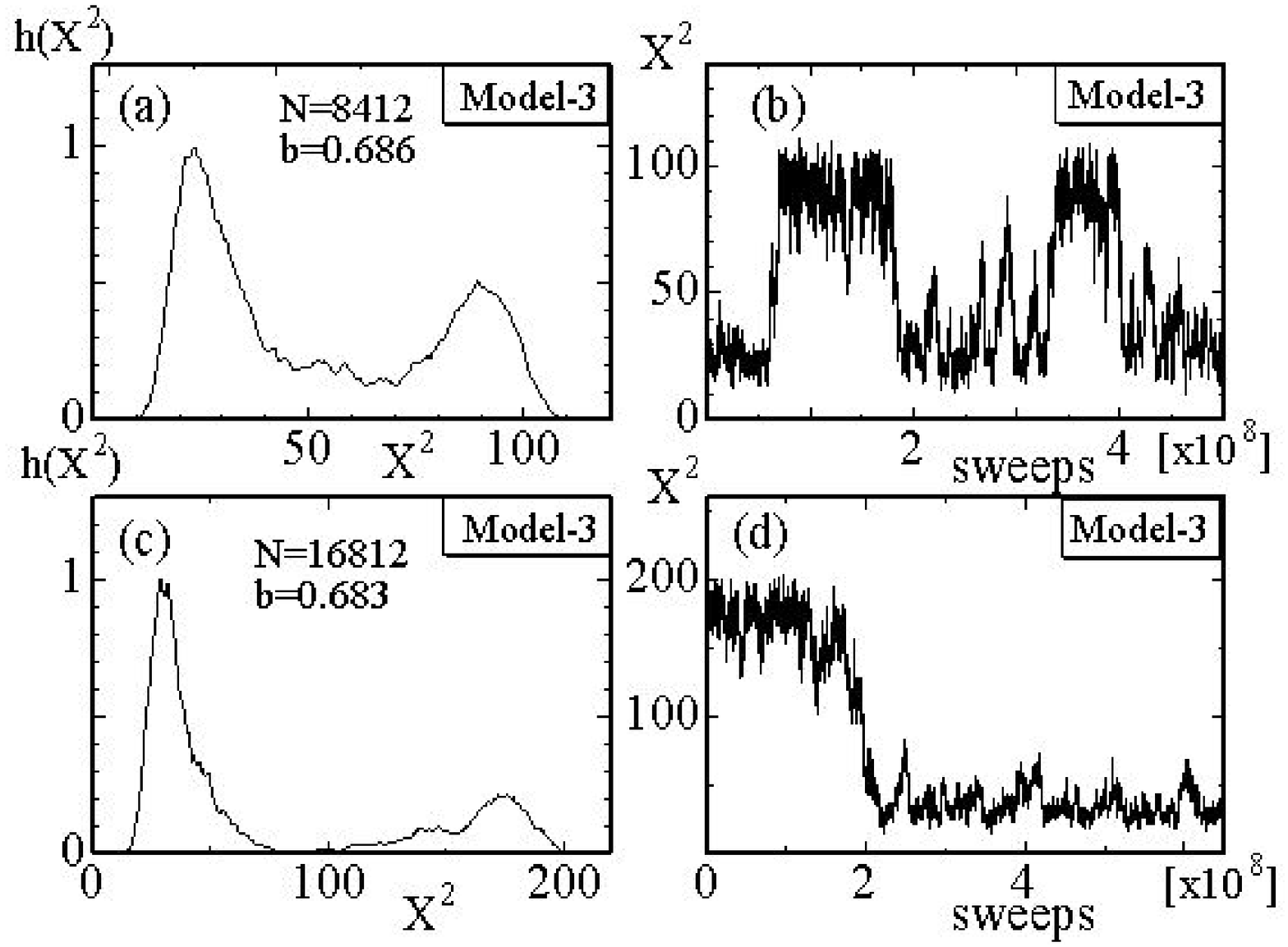}
 \caption{(a) The normalized histogram $h(X^2)$ of $X^2$ for model 3 on the $N\!=\!8412$ surface at $b\!=\!0.686$, and (b) the variation of $X^2$ against MCS, (c) $h(X^2)$ and (d) the variation of $X^2$ obtained on the $N\!=\!16812$ surface at $b\!=\!0.683$. }
\label{fig-4}
\end{figure}
Figure \ref{fig-4}(a) is a normalized histogram $h(X^2)$ of $X^2$ for model 3 on the surface of size $N\!=\!8412$ at the transition point $b\!=\!0.686$. The variation of $X^2$ against MCS is plotted in Fig. \ref{fig-4}(b). Figures  \ref{fig-4}(c) and \ref{fig-4}(d) are those obtained on the surface of $N\!=\!16812$ at $b\!=\!0.683$.  

In order to obtain the Hausdorff dimensions $H_{\rm smo}$ ($H_{\rm cru}$) in the smooth (crumpled) phase close to the transition point, the mean value of $X^2$ was obtained by averaging $X^2$ over a small region at each peak of $h(X^2)$. For model 1, we assumed the region: $4\!\leq\! X^2\!\leq\!10$ and $14\!\leq\! X^2\!\leq 19$ at $b\!=\!0.458$ on the $N\!=\!1442$ surface, $6\!\leq\! X^2\!\leq\!18$ and $20\!\leq\! X^2\!\leq 31$ at $b\!=\!0.456$ on the $N\!=\!2562$ surface, $7\!\leq\! X^2\!\leq\!25$ and $35\!\leq\! X^2\!\leq\!52$  at $b\!=\!0.4565$ on the $N\!=\!4142$ surface, and $8\!\leq\! X^2\!\leq\!30$ and $50\!\leq\! X^2\!\leq\!72$ at $b\!=\!0.455$ on the $N\!=\!6252$ surface. 

 For model 2, we assumed the region: $40\!\leq\! X^2\!\leq\!90$ and $100\!\leq\! X^2\!\leq 150$ at $b\!=\!0.115$ on the $N\!=\!2562$ surface, $50\!\leq\! X^2\!\leq\!150$ and $170\!\leq\! X^2\!\leq\!270$ at $b\!=\!0.1145$ on the $N\!=\!4842$ surface, $65\!\leq\! X^2\!\leq\!220$ and $230\!\leq\! X^2\!\leq\!380$ at $b\!=\!0.1144$ on the $N\!=\!6762$ surface, and $120\!\leq\! X^2\!\leq\!320$ and $360\!\leq\! X^2\!\leq\!560$ at $b\!=\!0.1144$ on the $N\!=\!10242$ surface.

 For model 3, we assumed the region: $8\!\leq\! X^2\!\leq\!21$ and $23\!\leq\! X^2\!\leq\!36$  at $b\!=\!0.692$ on the surface of $N\!=\!2562$,  $9\!\leq\! X^2\!\leq\!35$ and $39\!\leq\! X^2\!\leq\!64$ at $b\!=\!0.688$ on $N\!=\!4842$,  $11\!\leq\! X^2\!\leq\!49$ and $70\!\leq\! X^2\!\leq\!108$ at $b\!=\!0.686$ on $N\!=\!8412$, and  $16\!\leq\! X^2\!\leq\!55$ and $135\!\leq\! X^2\!\leq 205$ at $b\!=\!0.683$ on the surface of $N\!=\!16812$. 

\begin{figure}[hbt]
\centering
\includegraphics[width=12cm]{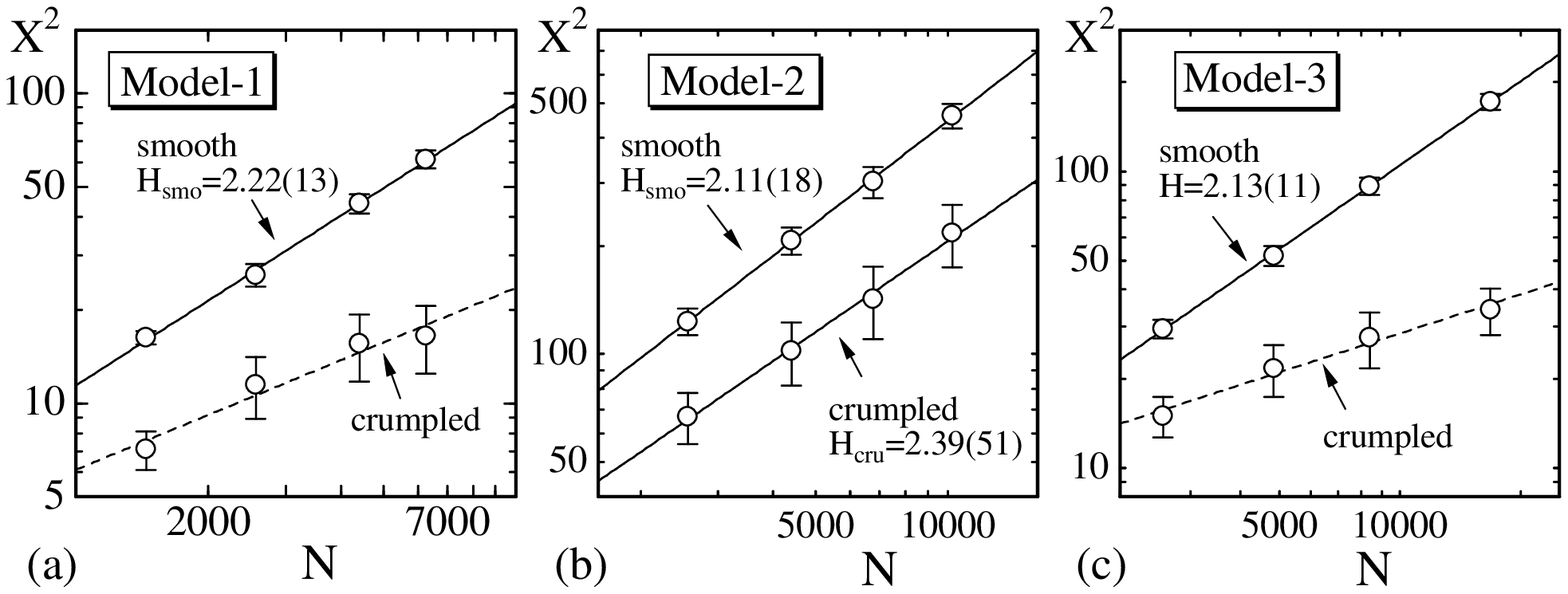}
 \caption{Log-log plots of $X^2$ against $N$ obtained in the smooth phase and in the crumpled phase close to the transition point for (a) model 1, (b) model 2, and  (c) model 3. The error bars on the data represent the standard deviations. }
\label{fig-5}
\end{figure}
Figures \ref{fig-5}(a),(b),(c)  are log-log plots of $X^2$ against $N$ for model 1, model 2, and model 3 respectively, where the values of $X^2$ are obtained by averaging $X^2$ in the regions described above.  Error bars on the data represent the standard deviations. The straight lines are drawn by fitting the data $X^2$ to
\begin{equation}
\label{Hausdorff-scale}
X^2 \sim N^{2/H}. 
\end{equation}
Then we have 
\begin{eqnarray}
\label{H-model}
&&H_{\rm smo}(1) = 2.22\pm 0.13  \quad ({\rm model \;1}),\nonumber \\
&&H_{\rm smo}(2) = 2.11\pm 0.18,  \quad H_{\rm cru}(2) = 2.39\pm 0.51 \quad ({\rm model \;2}),\\
&&H_{\rm smo}(3) = 2.13\pm 0.11 \quad ({\rm model \;3}).\nonumber
\end{eqnarray}
where $i$ in $H_{\rm smo, cru}(i)$ denotes the model $i(=1,2,3)$, and smo (cru) denotes the smooth (crumpled) phase. $H_{\rm smo}(1)$, $H_{\rm smo}(2)$, and   $H_{\rm smo}(3)$  in Eq. (\ref{H-model}) are comparable to the topological dimension $H\!=\!2$ within the errors. This data indicates that the surfaces are smooth in the smooth phase at the transition point for the three models.  

$H_{\rm cru}(2)$ in Eq.(\ref{H-model}) is less than the physical bound $H\!=\!3$. This implies that the surface is not completely crumpled but relatively smooth in the crumpled phase at the transition point of model 2. A similar phenomenon occurs for the standard tethered model \cite{KOIB-PRE-2005-1}. Note also that $H_{\rm cru}(2)$ in Eq. (\ref{H-model}) is comparable to the theoretical prediction $H=2.39(23)$ within the error,  which corresponds to the scaling exponent $\nu\!=\!0.84\!\pm\!0.04$ \cite{David-Wiese-PRL96} where $\nu\!=\!2/H$.

On the contrary, we got $H_{\rm cru}(1) \!=\! 3.27\!\pm\! 0.88$ in the crumpled phase at the transition point of model 1 by fitting the data to Eq. (\ref{Hausdorff-scale}). However, it seems that $H_{\rm cru}(1)$ is not well-defined because of the large error. The logarithmic divergence \cite{NELSON-SMMS2004,Gross-PLB1984,Duplantier-PLB1984,JK-PLB1984}; $X^2\!=\!c_0\!+\!c_1\log N$, is also expected for $X^2$ in the crumpled phase. In fact, the residual sum of squares RSS is ${\rm RSS}\!=\!0.424$ for the log-log fit of $X^2$ in the crumpled phase at the transition point, whereas the linear-log fit gives ${\rm RSS}\!=\!0.0906$ for these $X^2$, where ${\rm RSS}$ is defined by ${\rm RSS}=\sum [({\rm data}-{\rm fitting\; formula})/{\rm error}]^2$. The fact that the linear-log fit is better than the log-log fit indicates that $X^2$ scales according to the logarithmic divergence and that $H_{\rm cru}(1) \!=\! 3.27\!\pm\! 0.88$ is not well-defined. 

We also have $H_{\rm cru}(3)\!=\! 4.53\pm 1.23$ by fitting $X^2$ in the crumpled phase at the transition point of model 3. The dashed line in Fig. \ref{fig-5}(c) is drawn by using this $H_{\rm cru}(3)$, and indicates that $H_{\rm cru}(3)$ is ill-defined for the same reason as for $H_{\rm cru}(1)$. In fact, we have ${\rm RSS}\!=\!0.00438$ for the linear-log fit and ${\rm RSS}\!=\!0.221$ for the log-log fit in Fig. \ref{fig-5}(c). Thus, the linear-log fit is better than the log-log fit for $X^2$ in the crumpled phase of model 3. This allowed us to conclude that $H_{\rm cru}(3)\!=\! 4.53\pm 1.23$ is an ill-defined value. 

We understand also that a choice of discretization of the Hamiltonian influences the scaling property of $X^2$ in the crumpled phase at the transition point. 

\begin{figure}[hbt]
\centering
\includegraphics[width=8.5cm]{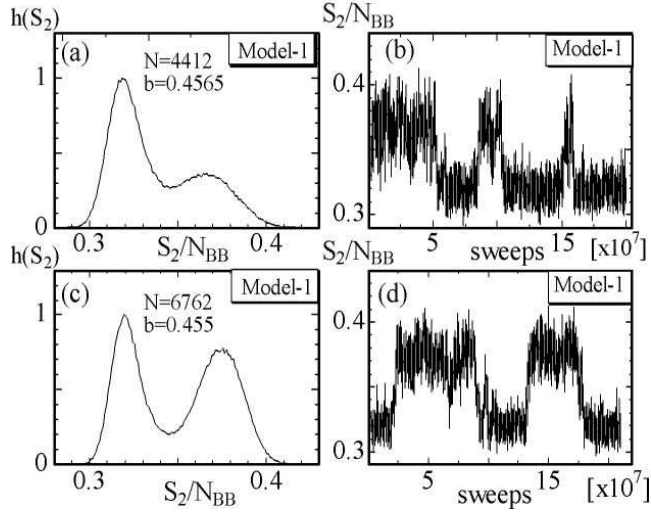}
 \caption{(a) The normalized histogram $h(S_2)$ of $S_2/N_{BB}$, and (b) the variation of $S_2/N_{BB}$ against MCS, on the $N\!=\!4412$ surface at $b\!=\!0.4565$,  (c) $h(S_2)$ and (d) the variation of $S_2/N_{BB}$ on the $N\!=\!6252$ surface at $b\!=\!0.455$. $N_{BB}\!=\!2N_B$. }
\label{fig-6}
\end{figure}
Figure \ref{fig-6}(a) is a normalized histogram $h(S_2)$ of $S_2/N_{BB}$ obtained on the surface of size $N\!=\!4412$ at the transition point $b\!=\!0.4565$, where $N_{BB}$ denotes $N_{BB}\!=\!2N_B$, and $N_B$ is the total number of bonds given by $N_B\!=\!3N-6$. The bending energy $S_2$ of model 1 is divided by $N_{BB}$ so that $S_2/N_{BB}$ can be regarded as $1\!-\!\langle \cos \theta\rangle$. Two peaks can be seen clearly in $h(S_2)$. The variation of $S_2$ against MCS is plotted in Fig. \ref{fig-6}(b). Figures \ref{fig-6}(c) and \ref{fig-6}(d) show $h(S_2)$ and the variations of $S_2$ obtained on  the surface of $N\!=\!6252$ at $b\!=\!0.455$. We also find two distinct peaks in $h(S_2)$ in Figs. \ref{fig-6}(c). These two peaks in Figs. \ref{fig-6}(a)  and \ref{fig-6}(c) obviously show that model 1 undergoes a first-order phase transition. Note also that the discontinuous transition is independent of the lattice structure. It was reported in \cite{KOIB-PRE-2004-1} that the same model undergoes a discontinuous transition on Voronoi triangulated spherical surfaces, which are different from those used in this paper.

\begin{figure}[hbt]
\centering
\includegraphics[width=8.5cm]{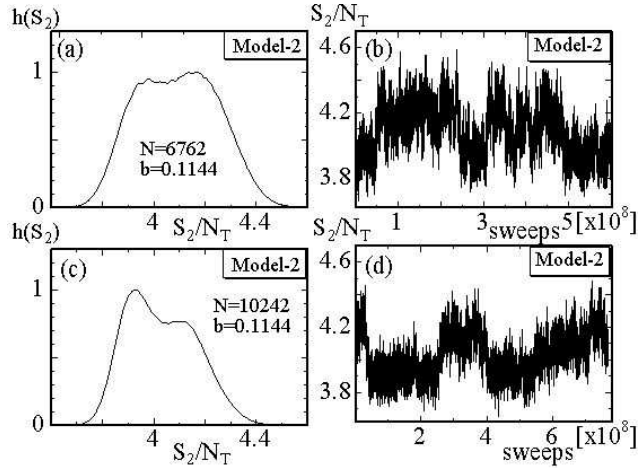}
 \caption{(a) $h(S_2)$ of $S_2/N_T$, and (b) the variation of $S_2/N_T$ against MCS, on the $N\!=\!6762$ surface at $b\!=\!0.1144$,  (c) $h(S_2)$ and (d) the variation of $S_2/N_T$ on the $N\!=\!10242$ surface at $b\!=\!0.1144$. $N_T\!=\!2N-4$. }
\label{fig-7}
\end{figure}
Figure \ref{fig-7}(a) is a normalized histogram $h(S_2)$ of $S_2/N_T$ obtained on the $N\!=\!6762$ surface at the transition point $b\!=\!0.1144$, where $N_T$ denotes the total number of triangles given by $N_T\!=\!2N-4$. The variation of $S_2$ against MCS is plotted in Fig. \ref{fig-7}(b). Figures  \ref{fig-7}(c) and \ref{fig-7}(d) show $h(S_2)$ and the variations of $S_2$ obtained on  the $N\!=\!10242$ surface at $b\!=\!0.1144$. The double peak structures of $h(S_2)$ in Figs. \ref{fig-7}(a)  and \ref{fig-7}(c) obviously show that model 2 undergoes a first-order phase transition. 

\begin{figure}[hbt]
\centering
\includegraphics[width=8.5cm]{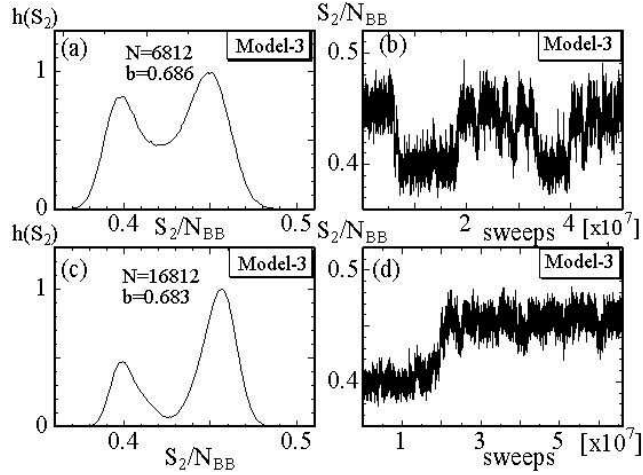}
 \caption{(a) $h(S_2)$ of $S_2/N_B$, and (b) the variation of $S_2/N_B$ against MCS, on the $N\!=\!8412$ surface at $b\!=\!0.686$,  (c) $h(S_2)$ and (d) the variation of $S_2/N_T$ on the $N\!=\!16812$ surface at $b\!=\!0.683$. $N_B\!=\!3N\!-\!6$. }
\label{fig-8}
\end{figure}
Figure \ref{fig-8}(a) is a normalized histogram $h(S_2)$ of $S_2/N_B$ obtained on the $N\!=\!8412$ surface at the transition point $b\!=\!0.686$. The variation of $S_2$ against MCS is plotted in Fig. \ref{fig-8}(b). Figures  \ref{fig-8}(c) and \ref{fig-8}(d) show $h(S_2)$ and the variations of $S_2$ obtained on  the $N\!=\!16822$ surface at $b\!=\!0.683$. The double peak structures in $h(S_2)$ in Figs. \ref{fig-8}(a)  and \ref{fig-8}(c) obviously show that model 3 undergoes a first-order phase transition. 

\begin{figure}[hbt]
\centering
\includegraphics[width=12cm]{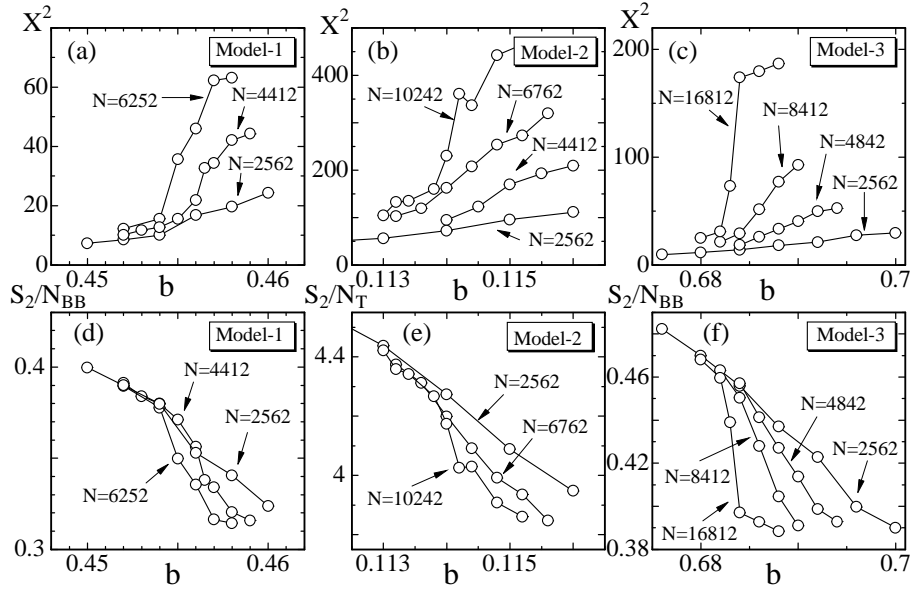}
 \caption{The mean square size $X^2$ against the bending rigidity $b$ of (a) model 1, (b) model 2, (c) model 3, and the bending energy (d) $S_2/N_{BB}$ against $b$ of model 1, (e) $S_2/N_T$ against $b$ of model 2,  (f) $S_2/N_{B}$ against $b$ of model 3.  $N_{BB}$ is twice that of $N_B$ the total number of bonds and $N_T$ is the total number of triangles. }
\label{fig-9}
\end{figure}

Figures \ref{fig-9}(a), (b), (c) show $X^2$ against $b$ obtained in model 1, model 2, and model 3, respectively. Figure  \ref{fig-9}(d)  shows the bending energy $S_2/N_{BB}$ against $b$ obtained in model 1. Figure \ref{fig-9}(e) shows $X^2$ and $S_2/N_T$ against $b$ obtained in model 2, and Fig. \ref{fig-9}(f) shows $X^2$ and $S_2/N_B$ against $b$ obtained in model 3. Note that the value for $S_2$ of model 2 is very different from that for model 1 or model 3. The reason for this is because $S_2/N_T$ can not be regarded as $1\!-\!\langle \cos \theta\rangle$, which can be demonstrated in Eq. (\ref{S1S2-model-2}). 

$X^2$, $S_2/N_{BB}$, and $S_2/N_T$ in  Figs. \ref{fig-9}(a), (b), (d), (e)  are almost smooth, and no discontinuous change is seen for intermediate bending rigidity $b$. The reason for those smooth behaviors is because the surface size is relatively small for model 1 and model 2. We expect that the discontinuous changes can be seen in those quantities obtained on large surfaces.  However, we consider that the sizes $N\!=\!6252$ and $N\!=\!10242$ are sufficiently large, because the discontinuous nature of the transition is obvious from the double peak in $h(S_2)$ in Figs. \ref{fig-6} and \ref{fig-7}. On the contrary, we see discontinuous changes in $X^2$, $S_2/N_{B}$ shown in  Figs. \ref{fig-9}(c), (f) for model 3. Surface size $N\!=\!16812$ seems sufficiently large for model 3 showing the discontinuity of $X^2$, $S_2/N_{B}$.

\begin{figure}[hbt]
\centering
\includegraphics[width=12cm]{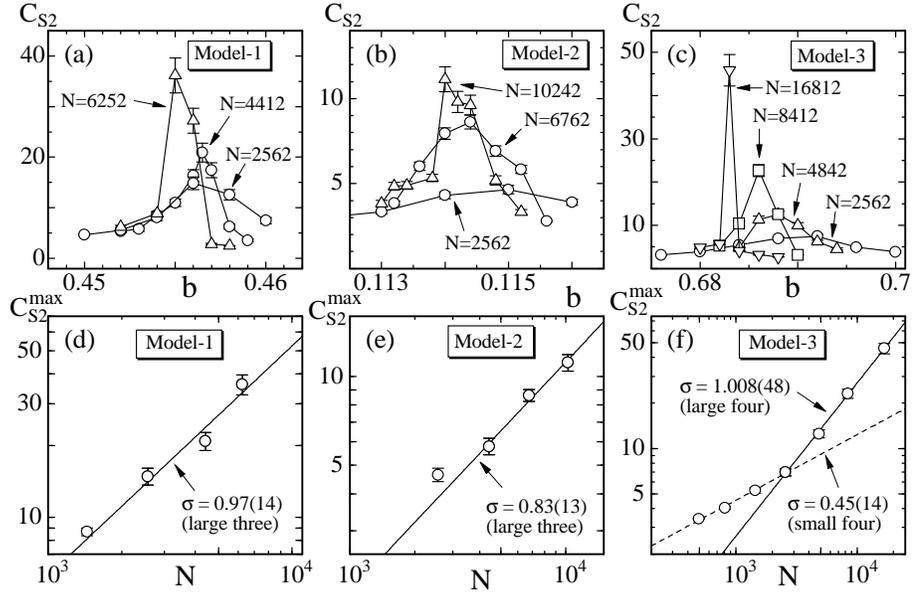}
 \caption{The specific heat $C_{S_2}$ against the bending rigidity $b$ of (a) model 1, (b) model 2, and (c) model 3.   Log-log plot of the peak values $C_{S_2}^{\rm max}$  against $N$ of (d) model 1,  (e) model 2, and (f) model 1. The straight lines in (d) and (e) are drawn by fitting the largest three $C_{S_2}^{\rm max}$ to Eq. (\ref{sigma-fitting}).  The solid (dashed) line in (f) is drawn by fitting the largest (smallest) four $C_{S_2}^{\rm max}$  to Eq. (\ref{sigma-fitting}).  }
\label{fig-10}
\end{figure}
The specific heat $C_{S_2}$ is defined by
\begin{equation}
\label{Spec-Heat}
C_{S_2} = {b^2\over N} \langle \; \left( S_2 - \langle S_2 \rangle\right)^2 \; \rangle.
\end{equation}
Figure \ref{fig-10}(a) shows $C_{S_2}$ of model 1 on the surfaces of size $N\!=\!6252$, $N\!=\!4412$, and $N\!=\!2562$. Sharp peaks in Fig. \ref{fig-10}(a) indicate that model 1 undergoes a discontinuous transition. Figure \ref{fig-10}(b) shows $C_{S_2}$ of model 2 on the surfaces of $N\!=\!10242$, $N\!=\!6762$, and $N\!=\!2562$, and Fig. \ref{fig-10}(c) shows $C_{S_2}$ of model 3 on the surfaces of $N\!=\!16812$, $N\!=\!8412$, $N\!=\!4842$, and $N\!=\!2562$.

Figure \ref{fig-10}(d) is  a log-log plot of the peak value $C_{S_2}^{\rm max}$ against $N$  including the results obtained on the $N\!=\!1442$ surface for model 1. The peaks $C_{S_2}^{\rm max}$  are plotted in Fig. \ref{fig-10}(e) and in  Fig. \ref{fig-10}(f) for model 2 and model 3, respectively. 

The straight lines in Figs. \ref{fig-10}(d), (e)  were drawn by fitting the largest three $C_{S_2}^{\rm max}$ to
\begin{equation}
\label{sigma-fitting}
C_{S_2}^{\rm max} \sim N^\sigma,
\end{equation}
where $\sigma$ is a critical exponent of the transition. The straight line denoted by {\it large four} ({\it small four}) in Fig. \ref{fig-10}(f) were drawn by fitting the largest (smallest) four $C_{S_2}^{\rm max}$  to Eq. (\ref{sigma-fitting}). 

Thus, we have
\begin{eqnarray}
\label{sigma-result}
\sigma_1=0.93\pm 0.28\qquad ({\rm model} \;1 ), \nonumber \\
\sigma_2=0.83\pm 0.13\qquad ({\rm model} \;2 ), \\
\sigma_3=1.008\pm 0.048,\quad ({\rm model} \;3 ),\nonumber
\end{eqnarray}
and $\sigma_3({\rm small})=0.45\pm 0.14\; (N\leq 2562) $ for model 3. The exponents $\sigma_1$, $\sigma_2$, $\sigma_3$ in Eq. (\ref{sigma-result}) indicates that the phase transition is of the first order in all three models. On the contrary, $\sigma_3({\rm small})\!=\!0.45(14)$ for $(N\leq 2562)$ indicates that model 3 appears to undergo a continuous transition on small-sized surfaces of $N \leq 2562$. The result $\sigma_3({\rm small})\!=\!0.45(14)$ is almost comparable to $\sigma\!=\!0.406(14)$ of the model on Voronoi triangulated spherical surfaces of size up to $N\!=\!1500$ in \cite{KOIB-PLA-2003-2}. Two different behaviors for $C_{S_2}^{\rm max}$ against $N$ shown in Fig. \ref{fig-10}(f) are consistent with the fact that two distinct peaks in $h(S_2)$ are observed only on large ($N\!\geq\! 4842$) surfaces.

\begin{figure}[hbt]
\centering
\includegraphics[width=12cm]{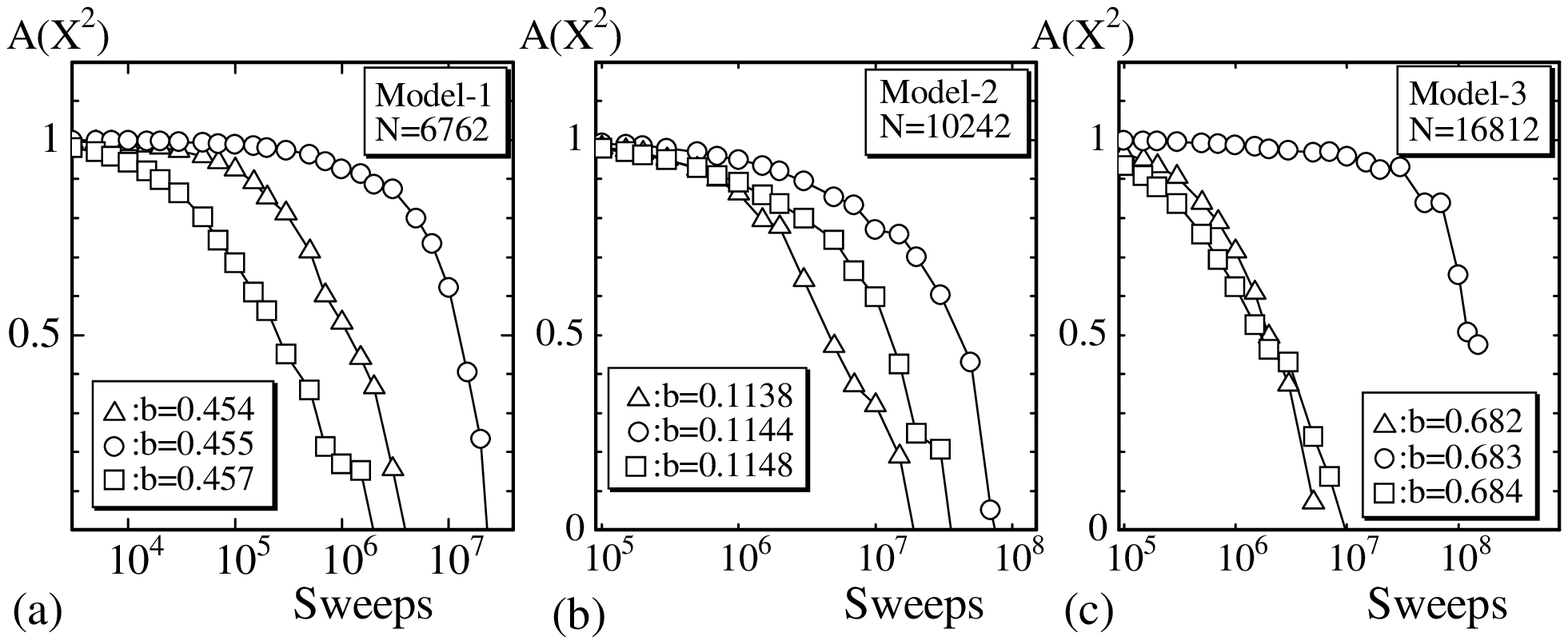}
 \caption{Autocorrelation coefficient $A(X^2)$ of $X^2$ obtained on (a) the model 1 surface of $N\!=\!6762$ at $b\!=\!0.454(\triangle)$,  $b\!=\!0.455(\bigcirc)$, and $b\!=\!0.457(\square)$, (b) the model 2 surface of $N\!=\!10242$ at $b\!=\!0.1138(\triangle)$,  $b\!=\!0.1144(\bigcirc)$, and $b\!=\!0.1148(\square)$, and (c) the model 3 surface $N\!=\!16812$ at $b\!=\!0.682(\triangle)$,  $b\!=\!0.683(\bigcirc)$, and $b\!=\!0.684(\square)$. }
\label{fig-11}
\end{figure}
Figures \ref{fig-11}(a),(b),(c) are the autocorrelation coefficient $A(X^2)$ of $X^2$ defined by
\begin{equation}
A(X^2)= \frac{\sum_i X^2(\tau_{i}) X^2(\tau_{i+1})} 
   {  \left[\sum_i  X^2(\tau_i)\right]^2 },\quad  \tau_{i+1} = \tau_i + n \times 500, \;(n=1,2,\cdots),  
\end{equation}
from which we can see the convergence speed of $X^2$ in the simulations. $A(X^2)$ in Fig. \ref{fig-11}(a) are obtained on the model 1 surface of $N\!=\!6762$ at $b\!=\!0.454$,  $b\!=\!0.455$, and $b\!=\!0.457$, and those in Fig. \ref{fig-11}(b) are obtained on the model 2 surface of $N\!=\!10242$ at $b\!=\!0.1138$,  $b\!=\!0.1144$, and $b\!=\!0.1148$. Figure \ref{fig-11}(c) shows $A(X^2)$ obtained on the model 3 surface $N\!=\!16812$ at $b\!=\!0.682$,  $b\!=\!0.683$, and $b\!=\!0.684$. In these figures, we find the phenomenon of critical slowing down, which reflects the transition. 

The horizontal axes in Figs. \ref{fig-11}(a),(b),(c) represent $500\times n\;(n\!=\!1,2,\cdots)$-MCS, which is a sampling-sweep between the samples $X^2(\tau_i)$ and $X^2(\tau_{i+1})$. The critical slowing down is clearly seen in the figures. The reason for this critical slowing down is because the volume of phase space ($\subseteq {\bf R}^3$) for $X$ at the transition point becomes larger than those at the crumpled phase and at the smooth phase. It is remarkable that the phase space volume in the smooth phase seems comparable to that in the crumpled phase, despite the fact that the surfaces in the smooth phase extend to large regions in ${\bf R}^3$. This can also be seen in a tethered model with intrinsic curvature on a sphere \cite{KOIB-EPJB-2004} and in a tethered model on a torus \cite{KOIB-PLA-2005-2}.

\begin{figure}[htb]
\centering
\includegraphics[width=9.5cm]{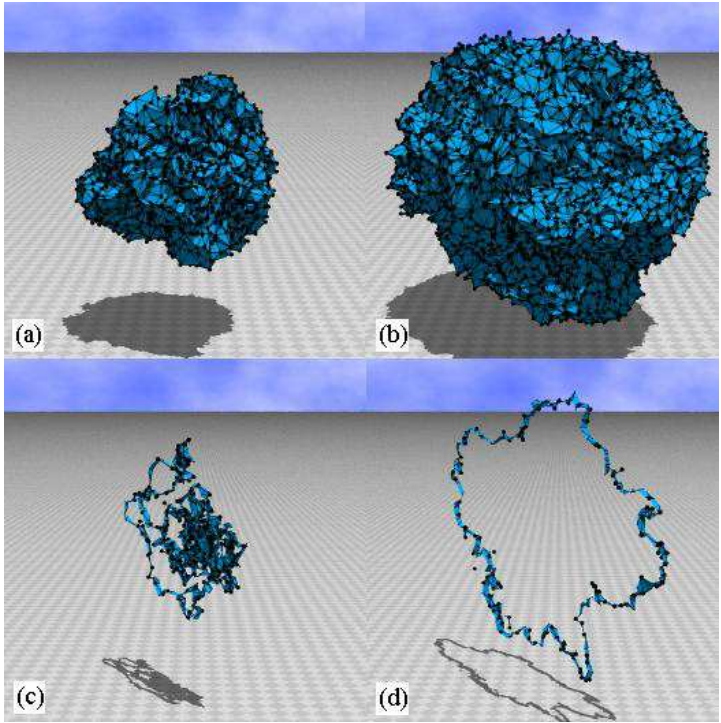}
 \caption{Snapshots of model 1 surfaces at (a) the crumpled phase and at (b) the smooth phase,  and  (c) the section of the surface in (a), and (d) the section of the surface in (b). The snapshots were obtained at the transition point $b\!=\!0.455$ on the surface of $N\!=\!6762$. }
\label{fig-12}
\end{figure}
Figure \ref{fig-12}(a) is a snapshot of model 1 surface of $N\!=\!6762$ in the crumpled phase at $b\!=\!0.455$, and Fig. \ref{fig-12}(b) is the one in the smooth phase at the same $b$. The sections of the surface shown in Figs. \ref{fig-12}(a) and \ref{fig-12}(b) are depicted in Figs. \ref{fig-12}(c) and \ref{fig-12}(d), respectively. Snapshots were drawn at the same scale.

\begin{figure}[htb]
\centering
\includegraphics[width=9.5cm]{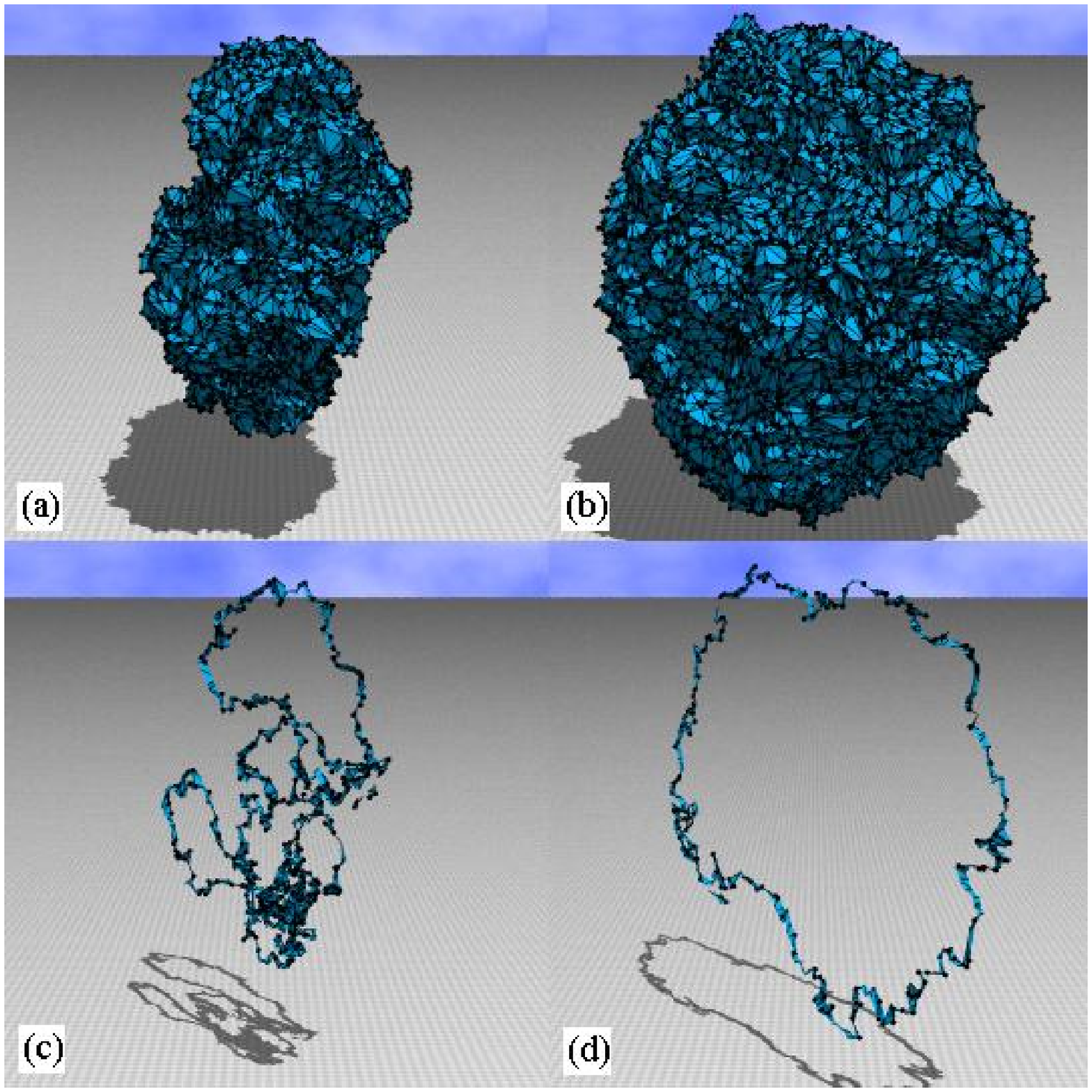}
 \caption{Snapshots of model 2 surfaces at (a) the crumpled phase and at (b) the smooth phase,  and  (c) the section of the surface in (a), and (d) the section of the surface in (b). The snapshots were obtained at the transition point $b\!=\!0.1144$ on the surface of $N\!=\!10242$. }
\label{fig-13}
\end{figure}
Figure \ref{fig-13}(a) is a snapshot of model 2 surface of $N\!=\!10242$ in the crumpled phase at $b\!=\!0.1144$, and Fig. \ref{fig-13}(b) is the one in the smooth phase at the same $b$. The sections of the surface shown in Figs. \ref{fig-13}(a) and \ref{fig-13}(b) are depicted in Figs. \ref{fig-13}(c) and \ref{fig-13}(d), respectively. Snapshots were drawn at the same scale.

\begin{figure}[htb]
\centering
\includegraphics[width=9.5cm]{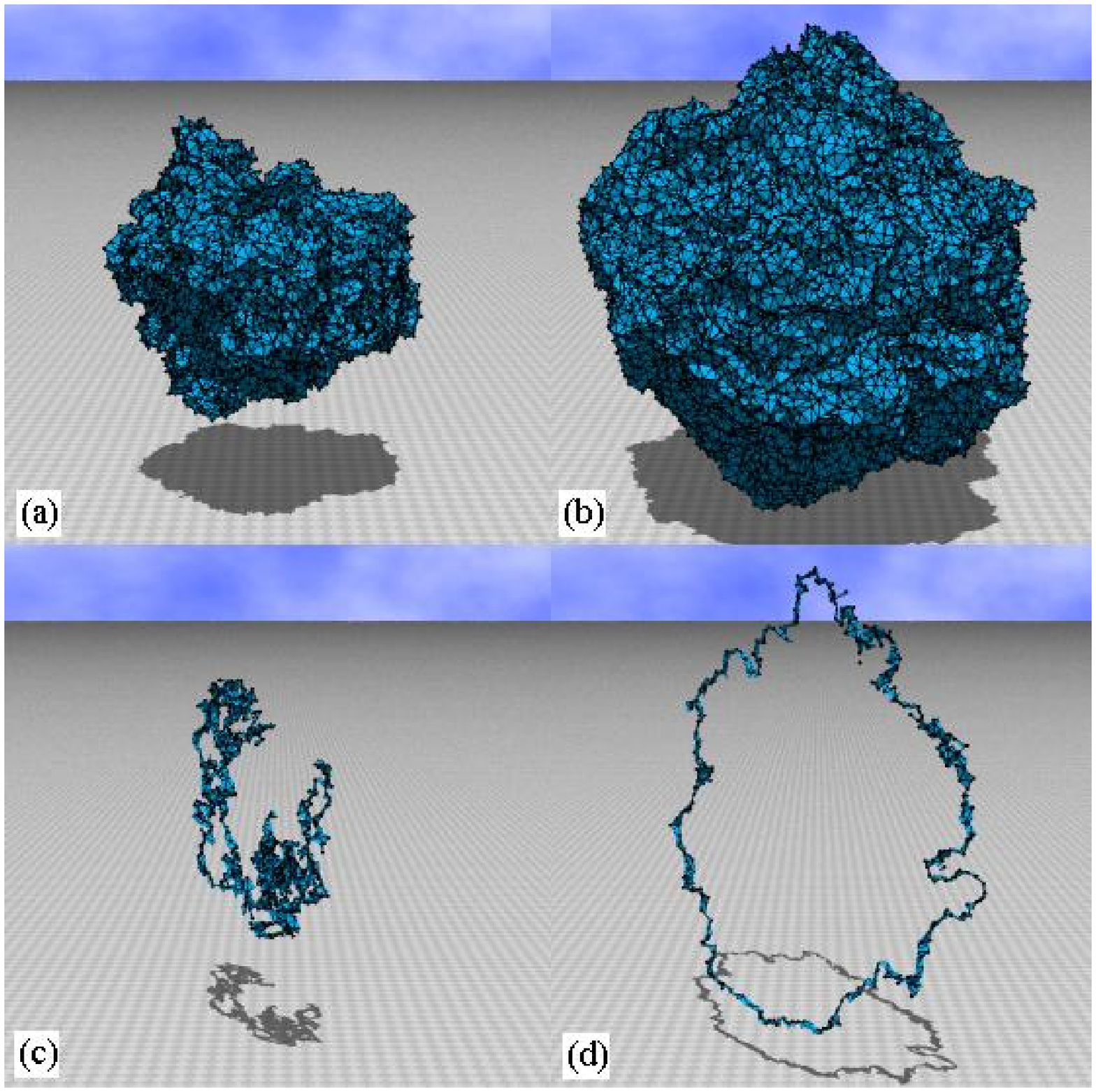}
 \caption{Snapshots of model 3 surfaces at (a) the crumpled phase and at (b) the smooth phase, and (c) the section of the surface in (a), and (d) the section of the surface in (b). The snapshots were obtained at the transition point $b\!=\!0.683$ on the surface of $N\!=\!16812$. }
\label{fig-14}
\end{figure}
Figure \ref{fig-14}(a) is a snapshot of the $N\!=\!16812$ surface in the crumpled phase at $b\!=\!0.683$, and Fig. \ref{fig-8}(b) is the one in the smooth phase at the same $b$. The mean square size is about $X^2\!=\!39$ and $X^2\!=\!178$ in Figs. \ref{fig-14}(a) and \ref{fig-14}(b), respectively. The sections of the surface in Figs. \ref{fig-14}(a) and \ref{fig-14}(b) are depicted in Figs. \ref{fig-14}(c) and \ref{fig-14}(d), respectively. 

Surfaces are rough at short scales even for the smooth phase shown in Figs. \ref{fig-12}(b), \ref{fig-13}(b), \ref{fig-14}(b), whereas they are smooth at the long range scale. Surfaces rough at short scales can also be seen deep in the smooth phase. Only a spherical monolayer surface is apparent in the smooth phase. None of non-spherical surfaces; such as oblong surfaces, linear ones, and branched-polymer like ones, were observed. The surfaces in the crumpled phase at the transition point are completely collapsed in both model 1 and model 3. These crumpled surfaces are in contrast to those in  model 2  and those in \cite{KOIB-PRE-2005-1}, where the crumpled state at the transition point seems not completely crumpled and seems to be characterized by a Hausdorff dimension less than the physical bound $H\!=\!3$. 

\section{Summary and conclusions}
This paper aimed to show that the first-order transition of the tethered surface model for biological membranes is independent of the discretization of the Hamiltonian. Three types of tethered surface models were investigated. The first model (model 1) is a discrete model of Helfrich and Polyakov-Kleinert,  the second (model 2) is that of Nambu-Goto, and the third (model 3) is a tensionless model of Helfrich and Polyakov-Kleinert. The first and the second models are closely related to each other in the context of the string model.

The canonical MC simulations were carried out on spherical surfaces of size up to $N\!=\!6252$ for model 1, $N\!=\!10242$ for model 2,  and $N\!=\!16812$ for model 3, where the lattices were constructed by dividing the icosahedron. A discrete form of the Hamiltonian of model 1 is a linear combination of the Gaussian term and a bending energy term, which is defined by using a normal vector of vertices. The Hamiltonian of model 2 contains an area term and a bending energy term, which makes the model well-defined. The Hamiltonian of model 3 contains a bending energy term and a hard-wall potential, which gives an upper bound on the bond length. The first-order transition of models 1, 2, and 3 were observed on lattices of size $N\!\geq\!2562$, $N\!\geq\!6762$, and $N\!\geq\!4842$, respectively. 

We have shown in this paper that the first-order crumpling transition is universal on spherical surfaces of the extrinsic curvature models for biological membranes. In fact, we have found that the first-order transition is independent of a choice of Hamiltonian of the tethered surface model, if it is discretized on a sphere. Therefore, the results obtained in this paper together with those in \cite{KOIB-PRE-2005-1} lead us to conclude that a first-order phase transition can be observed in a spherical tethered surface model whose Hamiltonian includes a bending energy term. It is also found that the Hausdorff dimension $H$ in the smooth phase at the transition point is almost identical to the topological dimension $H\!=\!2$ and is also independent of the discretization of the Hamiltonian. On the contrary, a choice of discretization of the Hamiltonian influences the scaling property of $X^2$ in the crumpled phase at the transition point. Nevertheless, we consider that the crumpled state of the model is not a phantom one but a real physical one, because the Hausdorff dimension of the crumpled phase in model 2 is $H_{\rm cru}(2) \!=\! 2.39(51)$, which is less than the physical bound and is also comparable to the result in \cite{KOIB-PRE-2005-1}.   

Finally, we emphasize that our results suggest that biological membranes can exhibit the crumpling transition. Experimental investigations on the crumpling transition are expected. Further numerical studies on the transition should also be performed on larger surfaces. 

This work is supported in part by a Grant-in-Aid of Scientific Research, No. 15560160. 



\end{document}